\documentclass[12pt]{article}
\usepackage{hyperref}
\usepackage{graphicx,psfrag,epsf}
\usepackage{setspace}
\usepackage{epsfig}
\usepackage{float}
\usepackage[centertags]{amsmath}
\usepackage{amsfonts}
\usepackage{url}
\usepackage{multirow}
\usepackage[round]{natbib}
\usepackage{bbm}     
\usepackage{dsfont}  

\newcommand{\indicator}[1]{\mathds{1}_{\left[ {#1} \right] }}

\begin{document}

\title{Mixture model with multiple allocations for clustering spatially correlated observations in the analysis of ChIP-Seq data}
\date{}
\author{Ranciati, S.$^{(*)(1)(2)}$, Viroli, C.$^{(1)}$, Wit, E.$^{(2)}$
\\[4pt]
\footnotesize (1) Department of Statistical Sciences \\[2pt]
\footnotesize University of Bologna, 40126 Bologna, Italy,\\[4pt]
\footnotesize email: saverio.ranciati2@unibo.it
\\[16pt]
\footnotesize (2) Johann Bernoulli Institute for Mathematics and Computer Sciences\\[2pt]
\footnotesize University of Groningen, 9747 AG Groningen, The Netherlands
}


\maketitle


\begin{abstract}
Model-based clustering is a technique widely used to group a collection of units into mutually exclusive groups. There are, however, situations in which an observation could in principle belong to more than one cluster. In the context of Next-Generation Sequencing (NGS) experiments, for example, the signal observed in the data might be produced by two (or more) different biological processes operating together and a gene could participate in both (or all) of them. We propose a novel approach to cluster NGS discrete data, coming from a ChIP-Seq experiment, with a mixture model, allowing each unit to belong potentially to more than one group: these multiple allocation clusters can be flexibly defined via a function combining the features of the original groups without introducing new parameters. The formulation naturally gives rise to a `zero-inflation group' in which values close to zero can be allocated, acting as a correction for the abundance of zeros that manifest in this type of data. We take into account the spatial dependency between observations, which is described through a latent Conditional Auto-Regressive process that can reflect different dependency patterns. We assess the performance of our model within a simulation environment and then we apply it to ChIP-seq real data.\\

\emph{KEYWORDS:} Heterogeneity; Model-based clustering; Spatial dependency; Multiple allocations.
\end{abstract}

\section{Introduction}
\label{sec1}
In the last 15 years, the development of parallel massively sequencing platforms for mapping the genome has completely revolutionized the way of studying gene expression patterns. These recent technologies called Next Generation Sequencing (NGS) allow to simultaneously investigate thousands of features within a single reliable and cost-effective experiment, thus representing a valid alternative to microarray experiments in enhancing our understanding of how genetic differences affect health and disease \citep{nag}. Indeed, these innovative platforms have been quickly applied to many genomic contexts giving rise to a large amount of available data in complex form. Data coming from such experiments are highly structured and their analysis has raised an imperative need for specific methodologies: technical or biological replicates can be observed in several experimental conditions \citep{Bao}, and the single observational units, such as genes or exons, are very likely characterized by spatial dependencies and relationships \citep{Mo}. Moreover, the abundance of a particular transcript is measured as a count and so the single data point is a discrete measurement.

\begin{figure}
\begin{center}
\centerline{\includegraphics[width=15.5cm]{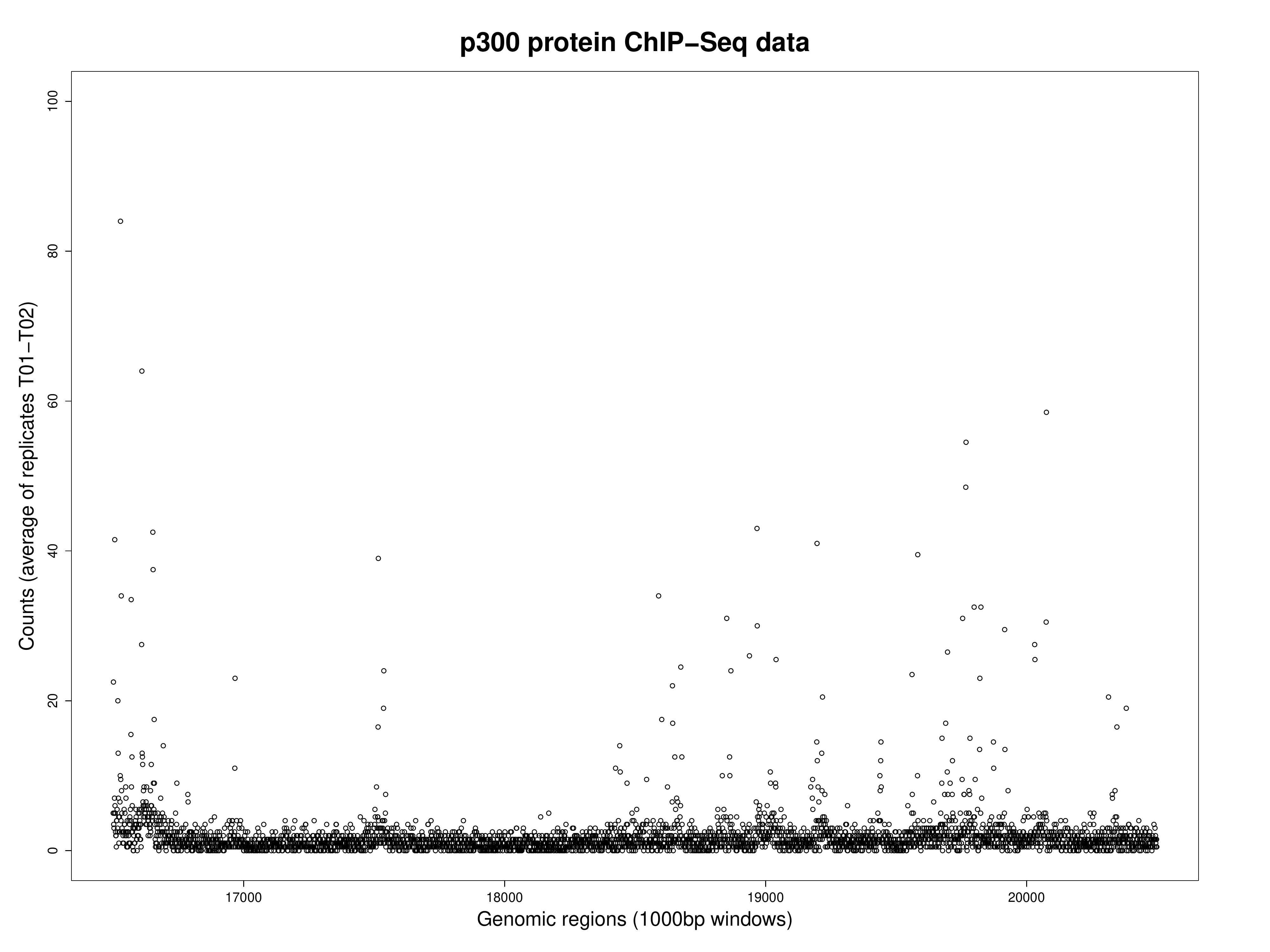}}
\caption{Summarized counts for 4000 bins from the chromosome 21. Regions on x-axis are 1000bp windows; averaged counts of the two replicates T01 and T02 reported on y-axis. \label{plot_zoom}}
\end{center}
\end{figure}

We consider in this paper the Chromatin ImmunoPrecipitation and sequencing (ChIP-Seq) data on proteins p300 and CBP analysed by \cite{Ramos2010}. In this experiment, two technical replicates are observed after 30 minutes from the initial interaction of the proteins with the human genome, with the aim of discovering the binding profile of these transcription factors. Figure \ref{plot_zoom} displays summarized counts for 1000 base pairs contiguous windows along chromosome 21. The plot shows segments with high read counts and segments where there is a uniformly low level of signal, thus suggesting a potential spatial effect.

In ChIP-Seq data like the one investigated (see \citet{kuan2011statistical} for a review), researchers usually associate the counts to two specific components: a background level, which accounts for the noise in the process and the inactivity of the regions with respect to protein binding; a signal level, that is described by a higher counts of sequenced DNA fragments, indicating that the protein is actually interacting with those specific genomic regions. From a statistical point of view, the problem of the detection of such biological processes can be addressed by introducing a mixture model with the aim of identifying groups of genomic regions that exhibit similar expression patterns. Typically, conventional model-based clustering methods perform classification of units into mutually exclusive partitions. However, looking at Figure \ref{plot_zoom}, it could be interesting to uncover components that may arise from the multiple overlapping action of the main aforementioned group processes. Multiple partitions can be obtained in (at least) two ways: (a) by fuzzy or `soft' clustering that is the assignment of a unit to the group with some posterior probabilities \citep[see, for instance, ][]{Bezdek, Heller2} or (b) by accounting for multiple allocations directly within a generative model. Our proposal employs this second perspective that explicitly assumes that genomic units are fully co-actors of multiple processes in a model based framework. The idea stems from earlier contributions aimed at discovering multiple clusters. \cite{Battle} introduced a probabilistic-based method to discover overlapping cellular processes and the associated gene regulation scheme. Given the complexity of a cellular system, they propose a decomposition of the observed continuous data matrix into layers representing biological processes and groups of co-regulated genes, allowing every unit to express itself in more than one activity layer and belong to multiple clusters. In \cite{Banerjee} and \cite{Fu}, the problem of multiple allocation is solved within a model based clustering strategy, where the distribution of the groups is extended generalizing the Gaussian probability distribution used in \cite{Battle} to the case of exponential family distributions. The main idea of such approach known as `Model-Based Overlapping Clustering' is to re-parameterize the density of the overlapped clusters as the product of some primary densities that, being members of the exponential family, still result in the same parametric family.
\cite{Heller1} extended these models by employing a nonparametric Bayesian technique to infer the number of groups in their overlapping clusters model, while maintaining the characterization of the mixture densities as members of the exponential family. More recently, \cite{Zhang} proposed the `epistatic model based clustering' for the analysis of microarray data. In this approach, a more explicit description of the mixed component densities in terms of Gaussians is given; different interactions between the parameters of the primary groups are investigated but the order of the interactions between these original clusters and the overlapped counterparts is practically limited to the second order.

The aim of this work is to define a general Multiple Allocation Mixture (MAM) model for analyzing the ChIP-Seq data. The peculiar features of these experimental data demand for specific treatment. First, their discrete nature and a marked overdispersion require a flexible count distribution such as the Negative Binomial, that however does not generally belong to the exponential family, unless its dispersion parameter is known and fixed. To this aim we generalize the model-based overlapping clustering to arbitrary parametric probabilistic functions. In addition, as shown in Figure \ref{plot_zoom}, ChIP-Seq data are characterized by the inflation of non-structural zeros. These aspects are naturally taken into account by the proposed model, where each component of the multiple mixture corresponds to a primary or to an overlapping cluster distributed as Negative Binomials with parameters that are function of the primary parameters only.
A further important aspect that emerges from Figure \ref{plot_zoom} is that the protein interactions with DNA are spatially correlated. We will show that the model can be easily extended in order to account for the spatial linkage among the genes, via a Conditional Auto-Regressive (CAR) process.

In what follows, we will present our proposal in three gradual steps in order to sequentially address these issues. First, in Section 2, the general MAM model will be presented and then we will adapt the model to the NGS data. We will illustrate how to extend this approach in order to model spatial dependent observations. In Section 3, a simulation study aimed at investigating the flexibility and effectiveness of the proposal is presented. Furthermore, we use the newly developed model to study the genome wide binding of the protein p300 in order to investigate its transcriptional regulatory function. In Section 4, conclusions are discussed.

\section{Methods}
\label{sec2}

\subsection{Model-based clustering with mixture model}
\label{model_based}

Finite mixture models have been receiving a wide interest in the statistical literature as a tool for performing model based clustering and density estimation \citep[see][]{Banfield1993,McLachlan2000,Fraley2002}.

Let $\boldsymbol{y}_j$ ($j=1,\dots,p$) be an observed vector of values that we want to classify in some unknown groups.  The conventional model based clustering model assumes
\begin{eqnarray}\label{mod1}
f(\boldsymbol{y}_j|\Theta)=\sum_{i=1}^k\pi_i f(\boldsymbol{y}_j|\boldsymbol{\theta}_i),
\end{eqnarray}
where $f(\boldsymbol{y}_j|\boldsymbol{\theta}_i)$ are the component densities with parameters $\theta_i$ and $\pi_i$ are the prior probabilities to belong to each component, satisfying $\pi_i>0$ and $\sum_{i=1}^k\pi_i =1$. According to this model, a sample of $p$ observations arises from some underlying populations of unknown proportions and the purpose is to decompose the sample into its mixture components, where each component corresponds to a cluster. In doing so, the data are partitioned into mutually exclusive $k$ groups. This is achieved by introducing a latent variable, say $z_{ji}$, which allocate each observation $j$ to the component $i$. More precisely, $\boldsymbol{z}_{j}$ is a vector of length $k$ that takes the value 1 in correspondence of the cluster assignment, and 0 elsewhere, so that $\sum_{i=1}^k z_{ji}=1$. According to the maximum a posteriori probability (MAP) rule, the partition is then obtained by assigning subjects to their most likely class according to the posterior probabilities of $z$ given the data:
\begin{eqnarray*}
f(z_{ji}|\boldsymbol{y}_j;\Theta)=\frac{\pi_if(\boldsymbol{y}_j|z_{ji};\boldsymbol{\theta}_i)}
{\sum_{i'=1}^k\pi_{i'}f(\boldsymbol{y}_j|z_{ji'};\boldsymbol{\theta}_{i'})}.
\end{eqnarray*}
In this sense the classification produced by a mixture model is `hard' (because a unit is allocated to the mixture component with the maximum posterior probability of belonging to) but in principles it could be `soft' by assigning each cluster a weight that equals its posterior probability as in the partial membership model \citep{Heller2}. However, a soft assignment perspective does not mitigate the limitation of the model based classification, that results when data points may simultaneously belong to multiple clusters. In such situations, a change in the generative model is required, by explicitly assuming that the allocation vector $\boldsymbol{z}_{j}$ may contain several - and not just a single - ones.

 \subsection{Multiple Allocation Mixture model}
\label{MAM}
In order to construct a new generative model for multiple components, we define $k$ in Eq. (\ref{mod1}) as the number of primary groups which are not mutually exclusive. For such a reason, we assume the prior probabilities of each primary group satisfying the constraint $\pi_i>0$ but not necessarily summing up to one,  $\sum_{i=1}^k\pi_i\neq 1$.

The total number of possible single and multiple allocation clusters is $k^*=2^{k}$ and $\sum_{i=1}^kz_{ji}=n_j$ is the multiplicity of the cluster membership for the unit $j$th.  When $n_j=1$ we have a single group allocation, otherwise we have multiple group allocations. More precisely, if $n_j=2$ the unit $j$th belongs to two groups simultaneously. These two groups altogether may be thought of as a new \emph{secondary} group. If $n_j=3$ the unit belongs to three groups that jointly define a \emph{tertiary} group and so forth. When $n_j=0$, the unit is assigned to what we call a `outward cluster': this group collects observations that ideally do not belong to any clusters, and for this reason their distribution might be described by peculiar parameters depending on the empirical context. For instance, in many applications it could represent a group of outliers or noisy observations, characterized by high variance. The definition and existence of the `outward cluster' is particularly relevant for the analysis of ChIP-Seq data, where the clusters are interpretable as biological processes. A gene that does not take part to any biological processes will have extremely low values (close to zero or zero). Thus, the  outward cluster has the purpose to describe the group of `inactive genes' and, in so doing, it acts as a zero-inflation adjustment for the model.

For $k$ fixed, let $U$ be a connection matrix of dimension $2^{k} \times k$, with elements $u_{hi} \in \{0,1\}$, containing all the possible assignment configurations. For instance, for $k=3$:
 \begin{eqnarray}\label{con.mat}
U= \left(
   \begin{array}{ccc}
     0 & 0 & 0 \\
     1 & 0 & 0 \\
     0 & 1 & 0 \\
     0 & 0 & 1 \\
     1 & 1 & 0 \\
     0 & 1 & 1 \\
     1 & 0 & 1 \\
     1 & 1 & 1 \\
   \end{array}
 \right).
\end{eqnarray}
In this case we define the prior probability to belong to a general (single or multiple allocation) group
\begin{equation}
\label{pi_star}
\pi^*_h=\prod_{i=1}^k \pi_i^{u_{hi}} (1-\pi_i)^{1-u_{hi}}
\end{equation}
with $h=1,\ldots,k^*$. Notice that these weights are constructed so that $\sum_{h=1}^{k^*}\pi^*_h=1$, and the definition comes naturally from the representation of clusters as sets in a sample space. 
 The Multiple Allocation Mixture model (MAM) can be defined as a reparameterization of Eq. \ref{mod1} in the new formulation
   \begin{eqnarray}\label{mod2}
f(\boldsymbol{y}_j|\Theta)=\sum_{h=1}^{k^*} \pi_h^* f(\boldsymbol{y}_j|\psi(\boldsymbol{u}_h,\boldsymbol{\theta}), \phi_h)
\end{eqnarray}
  where $\psi(\boldsymbol{u}_h,\boldsymbol{\theta})$ is a location parameter and $\phi_h$ is a nuisance parameter. More specifically,  $\psi(\boldsymbol{u}_h,\boldsymbol{\theta})$ is a function that depends on $\boldsymbol{u}_h$, which is the $h$-th row of $U$, and transforms the primary parameters $\boldsymbol{\theta}_i$ into the parameters of the multiple allocation components according to several possible schemes:
 \begin{itemize}
 \item \emph{additive model}: the parameters of the mixed groups could be the sum of the original parameters, that is $$\psi(\boldsymbol{u}_h,\boldsymbol{\theta})=\sum_{i=1}^k u_{hi}\boldsymbol{\theta}_i + \boldsymbol{\theta}_b\indicator{\sum_{i=1}^{k^*}u_{hi}=0},$$ where in the second term the indicator function is introduced to account for the parameters of the outward mixture component;
\item \emph{co-dominance of order 1}: $$\psi(\boldsymbol{u}_h,\boldsymbol{\theta})=\frac{\sum_{i=1}^k u_{hi}\boldsymbol{\theta}_i}{\sum_{i=1}^k u_{hi}}+ \boldsymbol{\theta}_b\indicator{\sum_{i=1}^{k^*}u_{hi}=0};$$
\item \emph{co-dominance of order 0}:
 $$\psi(\boldsymbol{u}_h,\boldsymbol{\theta})=\left(\prod_{i=1}^k \boldsymbol{\theta}_i^{u_{hi}}\right)^{1/\sum_{i=1}^k u_{hi}}+ \boldsymbol{\theta}_b\indicator{\sum_{i=1}^{k^*}u_{hi}=0}.$$
\end{itemize}
We will focus on the additive scheme through the rest of the manuscript. Finally, the nuisance parameters $\phi_h$ are taken h-specific because we aim to add more flexibility to the mixture model.

\subsection{Multiple Allocation Mixture model for ChIP-Seq Data}

Suppose we observe the ChIP-Seq counts of $p$ genes in $D$ biological conditions or replicates. We denote $Y_{jd}$ the random variable that expresses the read counts, say $y_{jd}$, mapped to gene $j$ ($j$=$1,...,p$), in sample $d$ with $d=1,\ldots,D$. Let $\boldsymbol{Y}_j$ be the random vector of length $D$ denoting the expression profile of a gene. Let $\boldsymbol{y}_j$ be the observed value. We assume that $Y_{jd}$ is distributed according to the Negative Binomial ($\mathcal{NB}$) distribution, given both the discrete nature of the observations and the flexibility of having a specific parameter used to model the overdispersion, which makes this distribution preferable over the Poisson. We further assume that, conditional on the group, the replicates are independent draws so that the mixture model in Eq. \ref{mod2} becomes:
   \begin{eqnarray}\label{mod3}
f(\boldsymbol{y}_j|\Theta)=\sum_{h=1}^{k^*} \pi_h^*
\prod_{d=1}^D \mathcal{NB}\left(y_{jd}|\psi(\boldsymbol{u}_h,\boldsymbol{\mu}_d),\phi_{dh}\right),
\end{eqnarray}
where $\phi_{dh}$ are specific dispersion parameters of the negative binomials and $\mu_{dh}^*=\psi(\boldsymbol{u}_h,\boldsymbol{\mu}_d)$ are the means defined in the extended space of multiple components. 

We allow dispersion parameters $\phi_{hd}$ to vary for each of the $k^*$ possible assignment and replicates because it is not directly clear a priori what is the most reasonable variance structure to define through a function for the multiple allocation clusters. This allows, nevertheless, for a certain degree of flexibility in describing the context-specific variability in the data.

With reference to the means $\mu_{dh}^*$ they are modelled as function of  primary $k$ means through the combination scheme $\psi$. More specifically, with reference to the connection matrix $U$ in Eq. \ref{con.mat}, $\mu_{d1}^*$ is the mean parameter in condition $d$ of the outward distribution; $\mu_{d2}^*$ is the mean parameter of the units that belong to the first primary group only and not to mixed groups, and so on, till $\mu_{d8}^*$ that is the mean of the units that belong simultaneously to the three components under condition $d$. Given the nature of the data we consider the `outward' group as the component devoted to describe the inactive genes and for this reason we fix its mean to a very low value, such as $\mu_{d1}^*=0.01$. The other mean parameters are obtained as combination scheme through $\psi$ of $k$ primary values that represent the means of the units that belong to the primary groups or related multiple groups. In order to construct an hierarchical form of the model, we define with $\boldsymbol{z}^*$ the allocation matrix in the augmented space given by $k^*$ as a $p \times k^*$ `multinomial' matrix. While each row of $\boldsymbol{z}$ can have multiple ones according to the assignment of the unit to multiple clusters, the rows of $\boldsymbol{z}^*$ only have a single one and the matrix $U$ allows for a connection between original parametrization allocation and the re-parametrized version. Given the whole set of parameters, say $\Theta$, the joint complete likelihood of our model is the following:
\begin{equation}\label{jcl}
P(\boldsymbol{y},\boldsymbol{z}^* | \Theta) = P(\boldsymbol{y} | \boldsymbol{z}^*,\Theta)\cdot P(\boldsymbol{z}^*|\Theta).
\end{equation}
Inference is carried out within a Bayesian framework. The posterior distribution of the parameters and the latent variables is:
\begin{equation}\label{postdens}
P(\boldsymbol{z}^*, \Theta, \boldsymbol{\xi} | \boldsymbol{y}) \propto P(\boldsymbol{y} | \boldsymbol{z}^*,\Theta)\cdot P(\boldsymbol{z}^*|\Theta) \cdot P(\Theta | \boldsymbol{\xi})
\end{equation}
with $P(\Theta | \boldsymbol{\xi})$ specifies priors for the mixture parameters and $\boldsymbol{\xi}$ as the vector containing the hyper-parameters. 
The posterior in Eq. \ref{postdens} can then be rewritten as:
\begin{eqnarray}\label{postNegBin}
\nonumber &P(\boldsymbol{z}^*,\boldsymbol{\pi}, \boldsymbol{\mu}, \boldsymbol{\phi}, \boldsymbol{\xi} | \boldsymbol{y}) \propto P(\boldsymbol{y} | \boldsymbol{z}^*,  \boldsymbol{\mu}, \boldsymbol{\phi}) P(\boldsymbol{z}^*| \boldsymbol{\pi}) P(\boldsymbol{\pi} | \boldsymbol{\xi}) P(\boldsymbol{\mu} | \boldsymbol{\xi}) P(\boldsymbol{\phi} | \boldsymbol{\xi}) = \\
& \nonumber  = \prod_{j=1}^{p} \prod_{d=1}^{D} \prod_{h=1}^{k^*} \left\{ \frac{\Gamma(\phi_{hd}+y_{jd})}{\Gamma(\phi_{hd})\Gamma(y_{jd}+1)} \left[ \frac{\phi_{hd}}{\phi_{hd}+\psi(\boldsymbol{u}_h, \boldsymbol{\mu}_d)} \right]^{\phi_{hd}} \left[ \frac{\psi(\boldsymbol{u}_h, \boldsymbol{\mu}_d)}{\phi_{hd}+\psi(\boldsymbol{u}_h, \boldsymbol{\mu}_d)} \right]^{y_{jd}} \right\}^{z^*_{jh}} \times \\
& \times \prod_{j=1}^{p} \prod_{h=1}^{k^*} \left( \prod_{i=1}^k\pi_i^{u_{hi}} (1-\pi_i)^{1-u_{hi}} \right)^{z^*_{jh}}   P(\boldsymbol{\pi} | \boldsymbol{\xi}) P(\boldsymbol{\mu} | \boldsymbol{\xi}) P(\boldsymbol{\phi} | \boldsymbol{\xi}),
\end{eqnarray}
where $ P(\boldsymbol{\pi} | \boldsymbol{\xi}), P(\boldsymbol{\mu} | \boldsymbol{\xi})$ and $ P(\boldsymbol{\phi} | \boldsymbol{\xi})$ are prior distributions for these quantities.
As prior distribution for the weights we assume a Beta distribution so that we get the following hierarchical structure:
\begin{eqnarray}
\label{hier}
\pi_i &\sim& \mbox{Beta}\left(1,1 \right) \nonumber  \\
P(\boldsymbol{z}_{jh}^* = 1 | \boldsymbol{\pi}^*) &=& \pi^*_{h} \\
\boldsymbol{y}_{jd} | \boldsymbol{z}^*_{jh} ; \boldsymbol{\Theta} & \sim & \mathcal{NB}(\psi(\boldsymbol{u}_h,\boldsymbol{\mu_d}); \boldsymbol{\phi}). \nonumber
\end{eqnarray}
For the other two parameters we select conjugate and flat priors. More precisely, for every $i$ and $h$, $\mu_{id} \sim \mbox{Gamma}(a_{\mu}, b_{\mu})$ and $\phi_{hd} \sim \mbox{Unif}(a_{\phi}, b_{\phi})$, where $(a_{\mu}, b_{\mu},a_{\phi}, b_{\phi})$ are elements of the vector $\boldsymbol{\xi}$ of hyperparameters. Given these chosen priors, full conditionals can be derived and a Gibbs sampling MCMC algorithm applied to estimate the parameters. At the implementation step of the algorithm we exploit the Gamma-Poisson mixture representation of a Negative Binomial distribution. We introduce a further latent variabile $s$, specific to each unit $j$ in each replicate/condition $d$, that has a Gamma density with shape and rate parameters equal to $\phi_{hd}$. It follows that:
\begin{eqnarray}
\nonumber &f(s_{jd}) = \mbox{Gamma}( \phi_{hd}, \phi_{hd} ) \\
\nonumber &p(y_{jd} | s_{jd} ) = \mbox{Pois}(\psi(\boldsymbol{u}_h, \boldsymbol{\mu}_d) s_{jd}) \\
\nonumber &p(y_{jd} ) = \int p(y_{jd} | s_{jd}) f(s_{jd}) ds = \mathcal{NB}(\psi(\boldsymbol{u}_h; \boldsymbol{\mu}_d),\phi_{hd})
\end{eqnarray}
which allows us to use a Gibbs sampler for the mean parameters $\mu_i$.

\subsection{Modelling spatial correlation with a CAR structure}

The units/observations (in our case, genes or genomic locations) are not independent but spatially correlated. We introduce the spatial correlation by allowing the primary weights $\pi_i $ to be $j$-varying and we denote them as $\pi_{ij}$. The spatial relationship is taken into account by allowing the weights of the mixture in Eq. \ref{mod2} to vary from one gene to another. The way we formulate it is inspired by the work proposed by \cite{Fern}. They first introduced $k$ independent $p$-dimensional latent variables with a Markov random-field distribution. The weights are then a non-linear function of these latent variables and spatial relationships are expressed in terms of neighborhood relationships. This could be restrictive because the correlation between two genes should decrease as their distance increases. We extend the approach by considering a Gaussian conditional auto-regressive model \citep{Pettitt}, where the distances are directly used to model correlations instead of dummies denoting the neighborhood condition. In a Bayesian framework, this is accomplished by introducing additional layers to the hierarchy formulation shown in Eq. \ref{hier}, with their own set of hyper-parameters. With reference to $\boldsymbol{\pi}$, we introduce the spatial latent vectors, denoted by $\boldsymbol{x}_1,\ldots, \boldsymbol{x}_i, \ldots, \boldsymbol{x}_k$, with $i=1,\ldots,k$: each $\boldsymbol{x}_i$ is a Gaussian conditional autoregressive model given by
\begin{equation}
\label{eqn:car}
f(\boldsymbol{x}_i)=\mathcal{N}(\boldsymbol{0},Q^{-1})
\end{equation}
where $Q$ is a precision matrix of order $p$ and $\gamma_{jj'}$ is some non-linear function of $\delta_{jj'}$, which are the distances between all the units. More specifically, 
$$Q=I_p + \Delta - \Gamma=\left\{
\begin{array}{ll}
	1+ \sum_{j'=1}^{p} \gamma_{jj'}  & \mbox{if $j$-th diagonal element} \\
	-\gamma_{jj'} & \mbox{elsewhere} \\
\end{array} \right.$$

After some algebraic steps it is possible to show that Eq. \ref{eqn:car} is equivalent to
\begin{eqnarray}\label{eqn:carbis}
f(\boldsymbol{x}_i)=
c \cdot \exp \left\{ -\frac{1}{2} \left[ \sum_{j=1}^{p} \sum_{j'=1}^{p} \gamma_{jj'} (x_{ij}-x_{ij'})^2 + \sum_{j=1}^{p} x_{ij}^{2} \right] \right\}
\end{eqnarray}
with $c$ the normalization constant
$$ c=(2\pi)^{-p/2} \prod_{j=1}^{p} (1+v_j)^{1/2}.$$
In the previous expression, $v_j$'s denote the eigenvalues of the spatial matrix $\Delta-\Gamma$. Given $\boldsymbol{x}_1,\ldots,\boldsymbol{x}_k$, the weights for location $j$ can be obtained via logistic formulation
$$ \pi_{ij}= \frac{\exp(x_{ij}/ \eta)}{1+\exp(x_{ij}/ \eta)} $$
where $\eta$ is a `shrink-or-stretch' tuning parameter to be estimated that provides a way to exaggerate the differences in units allocation among the clusters.

\subsection{Conditional Auto-Regressive Multiple Allocation Mixture (\emph{CAR-MAM})}

In order to account for spatial correlation between the units/observations (in our case, genes or genetic locations), we introduce another layer in the hierarchical structure of the model. Starting from Eq. \ref{jcl}, the updated joint complete likelihood is:
\begin{equation}
P(\boldsymbol{y},\boldsymbol{z}^*, \boldsymbol{x} | \boldsymbol{\mu}, \boldsymbol{\phi}, \eta, \boldsymbol{\xi}) = P(\boldsymbol{y} | \boldsymbol{z}^*,\boldsymbol{\mu}, \boldsymbol{\phi})  P(\boldsymbol{z}^*| \boldsymbol{x}, \eta)  P(\boldsymbol{x} | \boldsymbol{\xi})
\end{equation}
leading to the following posterior distribution
\begin{eqnarray}\label{postcar}
P(\boldsymbol{z}^*, \boldsymbol{x}, \boldsymbol{\mu}, \boldsymbol{\phi}, \eta |\boldsymbol{y}, \boldsymbol{\xi})  \propto P(\boldsymbol{y} | \boldsymbol{z}^*,\boldsymbol{\mu}, \boldsymbol{\phi})  P(\boldsymbol{z}^*| \eta, \boldsymbol{x})  P(\boldsymbol{x}| \boldsymbol{\xi}) P(\boldsymbol{\mu} | \boldsymbol{\xi}) P(\boldsymbol{\phi} | \boldsymbol{\xi})
\end{eqnarray}
where the vector $\boldsymbol{\xi}$ now also contains the hyper-parameters for the new latent layer and $P(\boldsymbol{y} | \boldsymbol{z}^*,\boldsymbol{\mu}, \boldsymbol{\phi})$ is defined as in Eq. (\ref{postNegBin}). More precisely, the complete latent structure in Eq. \ref{postcar} is equal to:
\begin{eqnarray*}
P(\boldsymbol{z}^*, \boldsymbol{x}| \eta) &= \prod_{j=1}^p \prod_{h=1}^{k^*} \left[ \prod_{i=1}^k\left( \frac{\exp(x_{ij}/\eta)}{1+\exp(x_{ij}/\eta)}  \right)^{u_{hi}}  \left(1- \frac{\exp(x_{ij}/\eta)}{1+\exp(x_{ij}/\eta)}  \right)^{1-u_{hi}}  \right]^{z_{jh}^*}  \times \\
\times & \prod_{i=1}^k c\exp \left\{ -\frac{1}{2} \left[  \sum_{j=1}^{p} \sum_{j'=1}^{p} \gamma_{jj'} (x_{ij}-x_{ij'})^2 + \sum_{j=1}^{p} x_{ij}^2 \right] \right\}
\end{eqnarray*}
where $c$ is the constant defined in Eq. \ref{eqn:carbis}.
As proposed in \cite{Fern}, we integrate out the latent allocation variable $\boldsymbol{z}$ when implementing the Metropolis sampler for $\boldsymbol{x}$ in order to employ the information carried by the data $\boldsymbol{y}$ and bring the two layers of the hierarchical structure of the model closer together.

\section{Results}
\label{sec3}

\subsection{Simulation study --- Multiple Allocation Mixture (\emph{MAM}) model}

We assess the performance of our model (\emph{MAM}) under different scenarios and we compare it with a classical non-overlapping components mixture (\emph{NegBinMix}). Data are generated from two independent Negative Binomial distributions ($D=2$) and $p=2000$ units, allowing for overlapping clusters with a number of groups $k=\{2,3\}$: in the augmented $k^*$ space this equals to represent a situation, as in a classical model-based clustering framework, where the actual number of groups ranges from $k^*=4$ to $k^*=8$. We explore three degrees of clustering between primary and outward/non-primary groups by selecting three scenarios which we call low, medium and high activation: this is achieved by setting all the $\pi$ equal to - respectively - $0.25$, $0.50$ and $0.75$. We run both algorithms (\emph{MAM} and \emph{NegBinMix}) for 10000 MCMC iterations and a 5000 burn-in window is selected. For every $\mu_{id}$, we choose hyperparameters of the Gamma prior distributions equal to $a_{\mu}=1$ and $b_{\mu}=0.001$; for every dispersion parameter $\phi_{hd}$, we select the ranges of the prior uniform distributions to be $[a_{\phi}=100,b_{\phi}=2000]$. Convergence is checked for every chain and we assign to the clusters according to the maximum a posteriori rule: we compute the posterior probabilities of the allocation vectors $\boldsymbol{z}^*_j$ given the data and, for every unit, we allocate to the component with the highest probability value (see also Section \ref{model_based}). We choose as an overall performance indicator the misclassification error rate (see Table \ref{tabNegBinMix}), that is, the average number of units not correctly allocated when compared to the known true membership. The posterior means for the parameters in the selected models (both the overlapping and non-overlapping mixtures) are consistent with the true values and do not show any substantial bias.

\begin{table}
\caption{Misclassification error rate (in percentage) for \emph{NegBinMix} and \emph{MAM} for the three scenarios of activation.}  \label{tabNegBinMix}
\begin{tabular}{ccccc}
\hline
\multirow{2}{*}{Number of clusters} & \multirow{2}{*}{Model} & \multicolumn{3}{|c|}{Degree of activation} \\
\cline{3-5}
& & $\pi_i = 0.25$ & $\pi_i = 0.50$ & $\pi_i = 0.75$ \\
\hline
$k=2$ & \emph{MAM} & \textbf{1.05} & 2.70 & 2.65 \\
(i.e. $k^*=4$) & \emph{NegBinMix} & 1.20 & \textbf{2.65} & \textbf{2.55} \\
\hline
$k=3$ & \emph{MAM}  & \textbf{0.95} & \textbf{3.05} & \textbf{5.10} \\
(i.e. $k^*=8$) & \emph{NegBinMix} & 11.00 & 25.05 & 9.15 \\ 
\hline
\\
\end{tabular}
\end{table}

We report the misclassification error rates for the estimated models in Table \ref{tabNegBinMix}: the percentage of misclassified units is always lower in the low activation scenario because most of the observations are allocated in the outward component, which has small variance and near zero (fixed) mean, thus simplifying the clustering task. As we can see in Table \ref{tabNegBinMix}, our model has comparable (sometimes better) classification rates, with respect to the conventional mixture of Negative Binomial, in simpler simulated dataset ($k=2$). When $k=3$, \emph{MAM} model always outperforms the compared mixture with noticeable improvements on the misclassification error rate.

\subsection{Simulation study --- \emph{MAM} with Conditional Autoregressive model (\emph{CAR-MAM})}

We simulate data from a mixture of Negative Binomials with multiple allocation and spatial information; we choose $k=2,3,4$ $(k^*=4,8,16)$ and for every number of groups we adopt two different spatial structures. In the former, the latent variable $\boldsymbol{x}$ is drawn from a CAR model using a reciprocal function $\gamma_{jj'}$ \citep[see][]{Pettitt}, which is an inverse function of the distances between uniformly drawn positions ($pos_j$, $j=1,\dots,p$). In the latter, a sine function is employed in the data generating process, in order to have stronger spatial relationships. More precisely, we assume that the spatial latent vectors, $\boldsymbol{x}_1,\ldots,\boldsymbol{x}_k$ are:
$$x_{ij}=sin\left(i  \pi  \frac{pos_j}{max\left\{pos_j\right\}_{j=1,\dots,p}}\right).$$ For each scenario we run three algorithms assuming $k$ is known: \emph{NegBinMix}, which is a mixture of Negative Binomials without any further specification, and our two proposed models \emph{MAM} and \emph{CAR-MAM} assuming that the conditional autoregressive structure is computed through a reciprocal function $\gamma_{jj'}$. We average the misclassification error rates across 100 independent datasets simulated for each scenario and we compare the results to assess the performance. Again, we cluster the units according to the maximum a posteriori rule and we measure the performance through the misclassification error, computed for each model. In Table \ref{tabCARMAM} and Table \ref{tabCARMAMsine} we summarize the results of 100 runs of both scenarios, where data were generated - respectively - with a reciprocal function for the \emph{CAR} part of the model or a sine function.

\begin{table}
\caption{Average misclassification error rate (in percentage) over 100 simulated datasets with standard errors in brackets; CAR structure with reciprocal function. \label{tabCARMAM}}
\begin{tabular}{cccc}
\hline
Number of clusters & \emph{CAR-MAM} & \emph{MAM} & \emph{NegBinMix} \\
\hline
$k=2$ $(k^*=4)$ & \textbf{26.58} (\emph{0.001}) & 26.87 (\emph{0.001}) & 30.40 (\emph{0.001}) \\
\hline
$k=3$ $(k^*=8)$ & \textbf{23.60} (\emph{0.001}) & 28.65 (\emph{0.009}) & 41.00 (\emph{0.002}) \\
\hline
$k=4$ $(k^*=16)$ & \textbf{36.32} (\emph{0.009}) & 49.84 (\emph{0.013}) & 60.39 (\emph{0.002}) \\
\hline
\\
\end{tabular}
\end{table}

As clear from the tables, we achieve improved accuracy in the clustering task with \emph{CAR-MAM} model over the simpler \emph{MAM} model and both of them perform better with respect to \emph{NegBinMix} (Table \ref{tabCARMAM}) even when the estimated conditional autoregressive structure is not the same as the one used in the data generating process (Table \ref{tabCARMAMsine}).

\begin{table}
\caption{Average misclassification error rate (in percentage) over 100 simulated datasets with standard errors in brackets; CAR structure with sine function.} \label{tabCARMAMsine}
\begin{tabular}{cccc}
\hline
Number of clusters & \emph{CAR-MAM} & \emph{MAM} & \emph{NegBinMix} \\
\hline
$k=2$ $(k^*=4)$ & \textbf{27.71} (\emph{0.002}) & 29.13 (\emph{0.001}) & 35.42 (\emph{0.001})\\
\hline
$k=3$ $(k^*=8)$ & \textbf{31.31} (\emph{0.006}) & 35.06 (\emph{0.011}) & 40.11 (\emph{0.002})\\
\hline
$k=4$ $(k^*=16)$ & \textbf{44.93} (\emph{0.005}) & 56.12 (\emph{0.007}) & 57.18 (\emph{0.003})\\
\hline
\\
\end{tabular}
\end{table}

\subsection{p300 protein binding ChIP-Seq experiment}

We apply our model to the data already discussed in the introduction and previously analysed by \cite{Ramos2010}. p300 is a transcription coactivator associated with many genes that are involved in multiple processes (i.e., differentiation, apoptosis, proliferation); also, the protein serves as a bridge for other transcription factors and it is involved in the transcription machinery of cells related to the development of cancer and other diseases. The data from p300 ChIP-Seq experiments consist of results from multiple interactions of the transcription coactivator with the chromatin in quiescent and stimulated cells, at different time points. We select two technical replicates (T01, T02) collected 30 minutes after the initial interaction between the chromatin and the protein p300 on a sequence of 1000 base-pairs windows describing the pre-processed raw counts across 4000 regions of the chromosome 21 (see Figure \ref{plot_zoom}). We analyze both replicates jointly and we summarize the data of this subsample by computing the average of the counts for the two different replicates T01 and T02. As we can see in the plot, the majority of the observations lie in a `band' of counts lower than 5, aggregated into segments that are spanned by smaller batch of regions exhibiting a higher count level, thus suggesting a spatial effect with respect to the protein binding process. Since in ChIP-Seq data researchers usually associate the counts to a background group and a signal group, we run the algorithm with $k=2$ ($k^*=4$) in order to capture the two aforementioned expected clusters and potentially a better characterization of them through our multiple allocation cluster, while simultaneously taking into account the spatial dependency between the genomic regions. We choose the geometric mean as our combination scheme $\psi(\cdot)$ to lessen the effect (on the multiple allocation cluster mean values) produced by the highest counts.
\begin{figure}
\includegraphics[width=15.5cm]{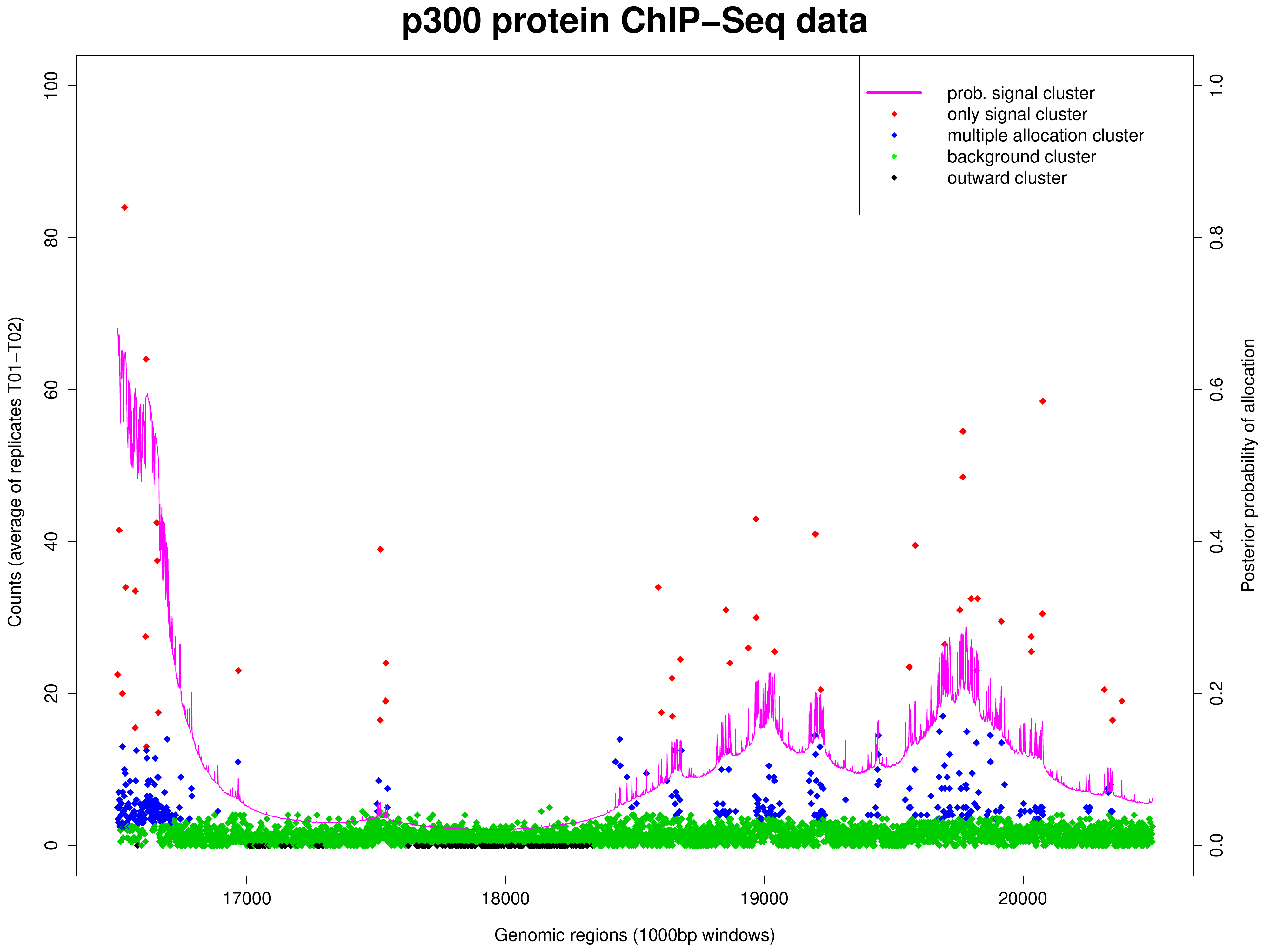}
\caption{Clustering result for the analyzed segment. Counts are reported as average of the two technical replicates T01 and T02; solid line is the posterior probability to be allocated to the signal group.\label{plot_results}}
\end{figure}
Out of the 4000 genomic regions, 159 are allocated in the outward cluster, which accounts for the zero-inflation in the data, represented by the observations around positions 17200 and 18000 (see Figure \ref{plot_results}). This group has fixed mean values equal to 0.01 for both replicates T01 and T02, while the dispersion parameters are estimated by the algorithm. The first primary cluster $i=1$, whose units are indicated by green dots in Figure \ref{plot_results}, has posterior means equal to $\boldsymbol{\mu}_1=(1.61, 1.07)$ and represents the background process of the protein binding: 3478 possibly inactive genomic regions, with very low counts, are allocated in this group. The second primary cluster, $i=2$, is representative of a signaling group of 48 genomic regions having a higher mean count level of $\boldsymbol{\mu}_2=(19.27,34.23)$ depicted with red dots in Figure \ref{plot_results}. The multiple allocation cluster in our analysis is interpretable as a group of 315 units involved in both the background and signal clusters: in this case, given that we are observing these counts after 30 minutes from the initial interaction of the protein with the strand of chromosome 21, the genomic regions allocated in this cluster could be thought as either being locations that were active immediately at the beginning and now not signaling anymore or locations only starting to interact with the p300 protein after 30 minutes. The mean values for this multiple allocation cluster are equal to $5.57$ and $6.06$ in the two replicates T01 and T02: the units belonging to it are shown as blue dots in Figure \ref{plot_results}. Finally, the posterior probability for each unit to be allocated in the signaling group is shown in Figure \ref{plot_results} as a magenta solid line. We can see from the plot that this probability is higher in those segments where genomic regions with higher counts are observed; moreover, the allocation weight of the signal component follows a spatial pattern, increasing and decreasing across the analyzed strand and thus accounting for the spatial effect occurring among the observations. For comparison, we also run the algorithm for a conventional four components Negative Binomial mixture model, fixing the location parameter of the first component to 0.01 as we do in our model.The result shows that only a total of three clusters are estimated, with one of them capturing the same background mean level shared by the fixed component, thus producing a blurred classification with respect to one of the two main processes of interest.

\section{Conclusions}
\label{sec4}

Motivated by the analysis of data coming from ChIP-Seq experiments, we proposed an extension of the conventional mixture model that allows for units to simultaneously belong to more than one group. As a by product of the model specification, an outward cluster has been introduced that can be used to describe specific features of the data such as zero-inflation, outliers and so forth. A spatial dependency layer among the units is encoded in the formulation by the means of a conditional autoregressive model, allowing for spatial information to aid the clustering task. We have compared our proposed model with a mixture of Negative Binomials to investigate its advantages with respect to the conventional approach: results on the simulated data show an increase in performance in terms of misclassification error rates. We applied our model to data previously analyzed by \cite{Ramos2010}: a promising richer description of the signal in the observations is found, calling for a potentially deeper biological investigation of those genomic regions associated with it. 
A delicate issue that deserves future investigation is the choice of $k$, which is a fundamental aspect because it identifies the number of primary clusters. Some well-known methods such as reversible jump \citep{GreenRJ} and birth-death process \citep{Stephens} do not allow an extensive exploration of the range of values for $k$ without incurring in a dramatic increase in computational cost. However, in some applications, this choice could be suggested by the empirical context such as in our case, where a number of primary clusters equal to two was reasonable because they represented background and signal groups.

\section*{Acknowledgments}

{\it Conflict of Interest}: None declared.

\begin{table}[!p]
\caption{Misclassification error rate (in percentage) for \emph{NegBinMix} and \emph{MAM} for the three scenarios of activation.}
\begin{center}
\begin{tabular}{|c|c|c|c|c|}
\hline
\multirow{2}{*}{Number of clusters} & \multirow{2}{*}{Model} & \multicolumn{3}{|c|}{Degree of activation} \\
\cline{3-5}
& & $\pi_i = 0.25$ & $\pi_i = 0.50$ & $\pi_i = 0.75$ \\
\hline
$k=2$ & \emph{MAM} & \textbf{1.05} & 2.70 & 2.65 \\
($k^*=4$) & \emph{NegBinMix} & 1.20 & \textbf{2.65} & \textbf{2.55} \\
\hline
$k=3$ & \emph{MAM}  & \textbf{0.95} & \textbf{3.05} & \textbf{5.10} \\
($k^*=8$) & \emph{NegBinMix} & 11.00 & 25.05 & 9.15 \\
\hline

\end{tabular}
\end{center}
\label{tabNegBinMix}
\end{table}%

\begin{table}[!p]
\caption{Average misclassification error rate (in percentage) over 100 simulated datasets with standard errors in brackets; CAR structure with reciprocal function.}
\begin{center}
\begin{tabular}{|c|c|c|c|}
\hline
Number of clusters & \emph{CAR-MAM} & \emph{MAM} & \emph{NegBinMix} \\
\hline
$k=2$ $(k^*=4)$ & 26.58 (\emph{0.001}) & 26.87 (\emph{0.001}) & 30.40 (\emph{0.001}) \\
\hline
$k=3$ $(k^*=8)$ & 23.60 (\emph{0.001}) & 28.65 (\emph{0.009}) & 41.00 (\emph{0.002}) \\
\hline
$k=4$ $(k^*=16)$ & 36.32 (\emph{0.009}) & 49.84 (\emph{0.013}) & 60.39 (\emph{0.002}) \\
\hline

\end{tabular}
\end{center}
\label{tabCARMAM}
\end{table}%

\begin{table}[!p]
\caption{Average misclassification error rate (in percentage) over 100 simulated datasets with standard errors in brackets; CAR structure with sine function.}
\begin{center}
\begin{tabular}{|c|c|c|c|}
\hline
Number of clusters & \emph{CAR-MAM} & \emph{MAM} & \emph{NegBinMix} \\
\hline
$k=2$ $(k^*=4)$ & 27.71 (\emph{0.002}) & 29.13 (\emph{0.001}) & 35.42 (\emph{0.001})\\
\hline
$k=3$ $(k^*=8)$ & 31.31 (\emph{0.006}) & 35.06 (\emph{0.011}) & 40.11 (\emph{0.002})\\
\hline
$k=4$ $(k^*=16)$ & 44.93 (\emph{0.005}) & 56.12 (\emph{0.007}) & 57.18 (\emph{0.003})\\
\hline

\end{tabular}
\end{center}
\label{tabCARMAMsine}
\end{table}%

\begin{figure}[!p]
\begin{center}
\includegraphics[width=12cm,keepaspectratio=TRUE]{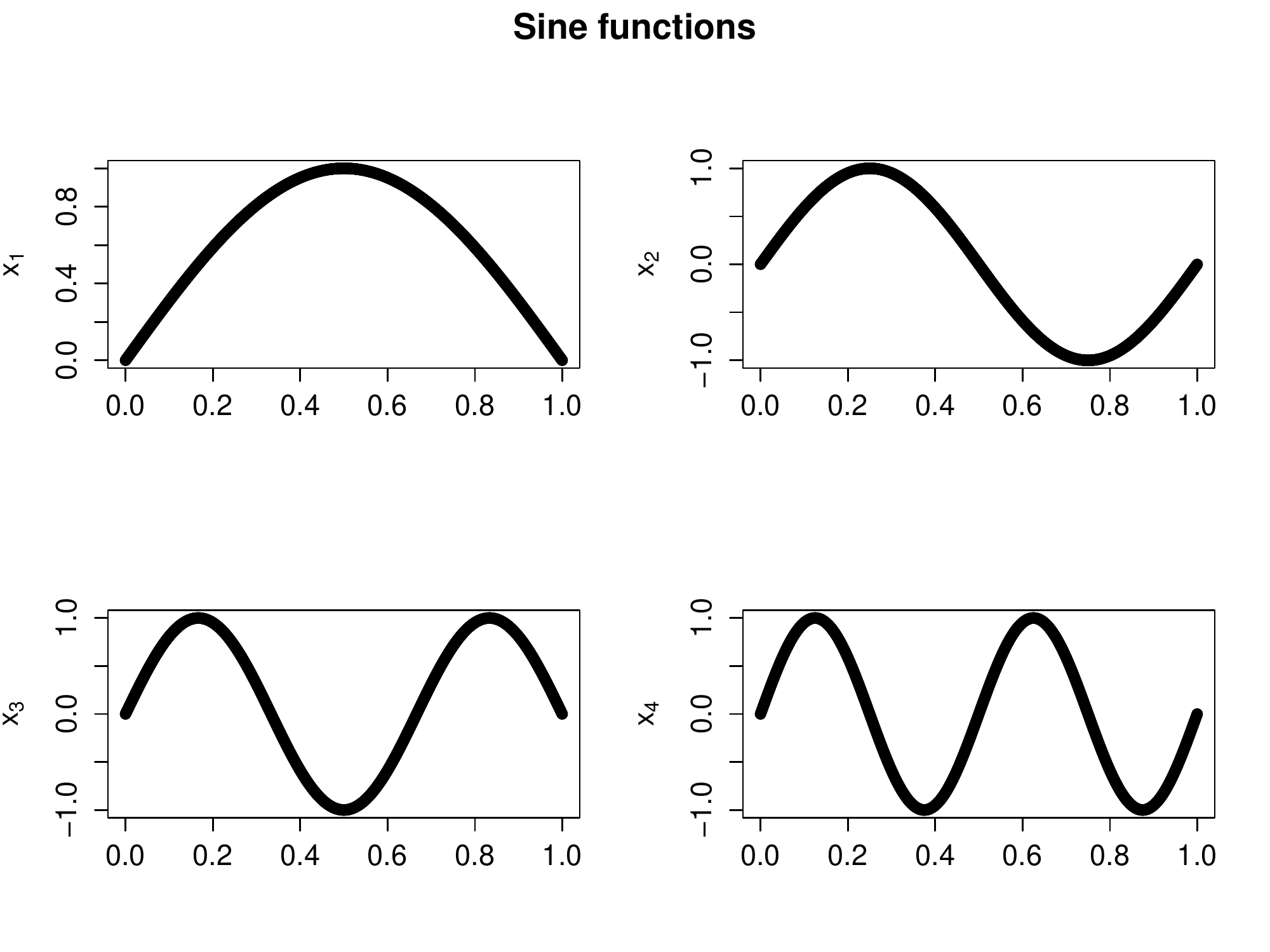}
\caption{Graphical visualization of the sine functions for $i=1$ up to $k=4$.}
\label{sine}
\end{center}
\end{figure}

\begin{figure}[!p]
\begin{center}
\includegraphics[width=15cm,keepaspectratio=TRUE]{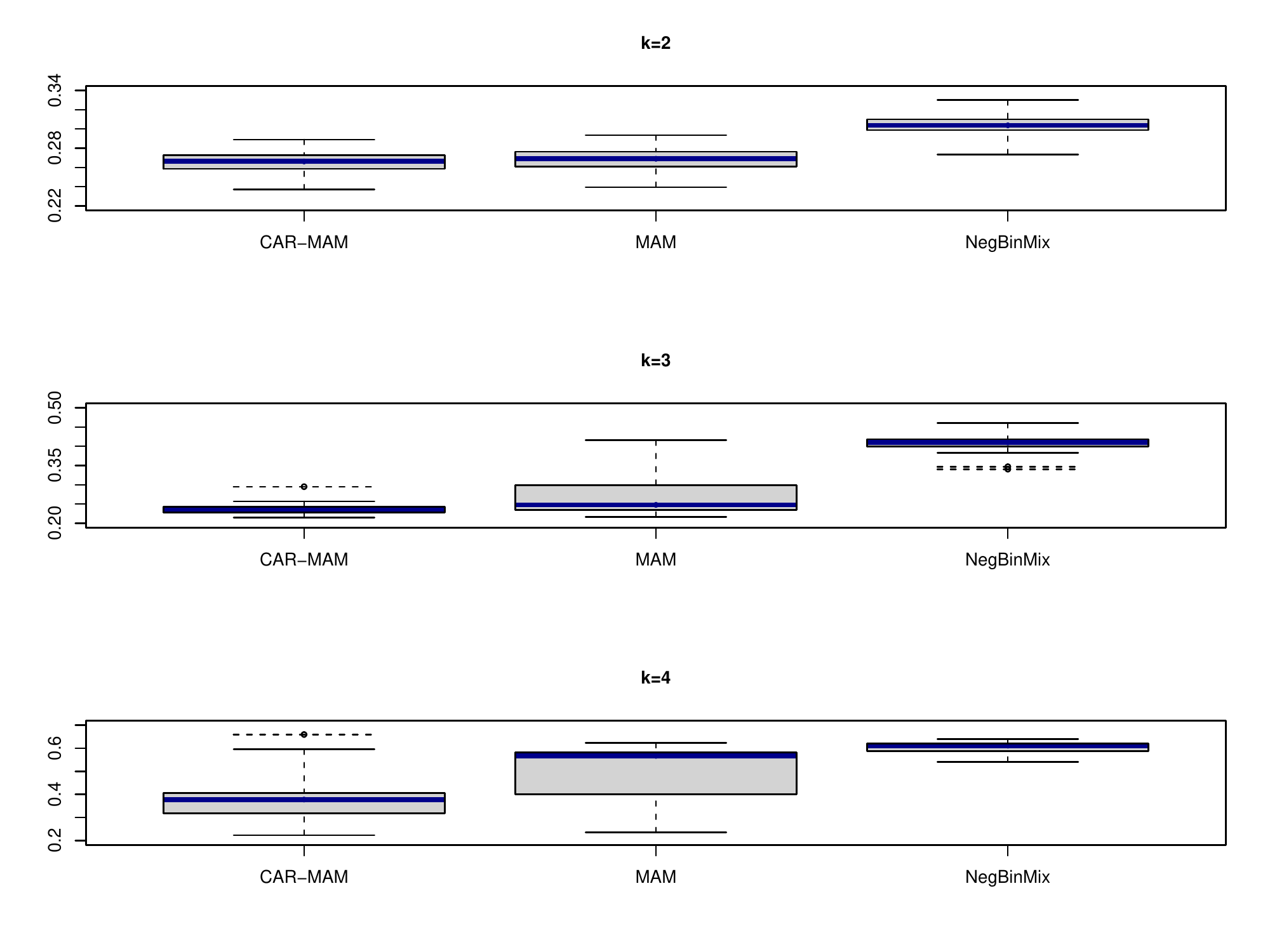}
\caption{Boxplots of misclassification errors over 100 datasets generated from CAR model using reciprocal function.}
\label{figCARMAM}
\end{center}
\end{figure}

\begin{figure}[!p]
\begin{center}
\includegraphics[width=15cm,keepaspectratio=TRUE]{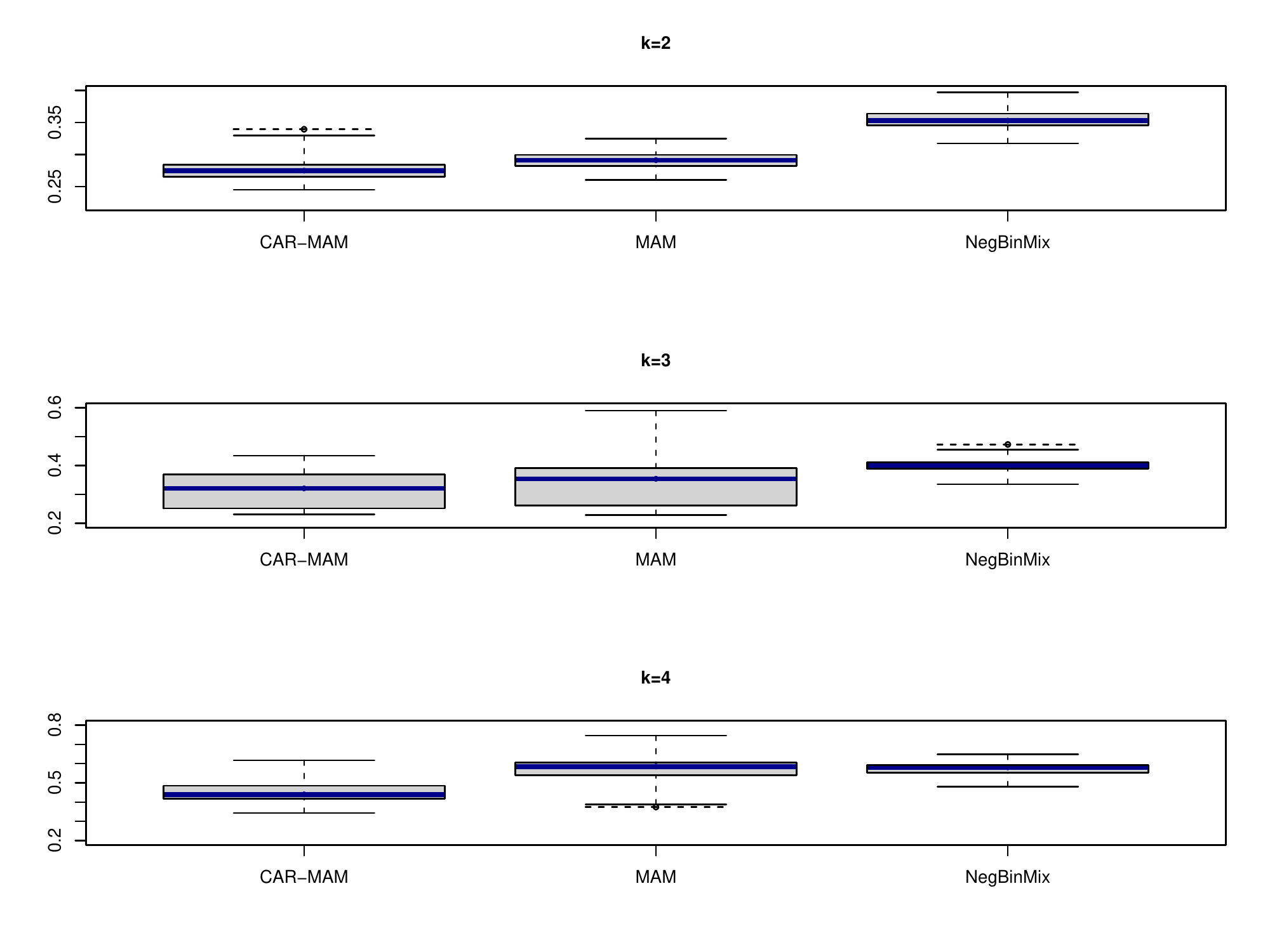}
\caption{Boxplots of misclassification errors over 100 datasets generated from CAR model using sine function.}
\label{figCARMAMsine}
\end{center}
\end{figure}

\end{document}